\documentclass[twocolumn,showpacs,preprintnumbers,amsmath,amssymb]{revtex4}
\usepackage{graphicx}% Include figure files
\usepackage{dcolumn}% Align table columns on decimal point
\usepackage{bm}% bold math
\newcommand{\SU }{\ensuremath{SU(3)}\,\,}
\newcommand{\Tc}{\ensuremath{\theta_{C}} }
\input epsf
\begin{document}

\preprint{APS/123-QED}
\title{Semileptonic Hyperon Decays and CKM Unitarity%
\footnote{To be published in Physical Review Letters.}\\}% Force line breaks with \\
\author{Nicola Cabibbo}
\affiliation{Department of Physics, University of Rome - La Sapienza\\
			and INFN, Sezione di Roma 1\\
Piazzale A. Moro 5, 00185 Rome, Italy\\
nicola.cabibbo@roma1.infn.it}

\author{Earl C. Swallow}
\affiliation{Department of Physics, Elmhurst College\\
Elmhurst, Illinois 60126 
and Enrico Fermi Institute, The University of Chicago, Chicago, Illinois\\ 
earls@elmhurst.edu}

\author{Roland Winston}
\affiliation{Division of Natural Sciences,\\
The University of California - Merced, Merced, California 95344\\
rwinston@ucmerced.edu}

%\author{Charlie Author}
 %\homepage{http://www.Second.institution.edu/~Charlie.Author}
%\affiliation{
%Second institution and/or address\\
%This line break forced% with \\}%
\date{\today}% It is always \today, today,
             %  but any date may be explicitly specified
%DRAFT DRAFT DRAFT DRAFT DRAFT DRAFT DRAFT DRAFT DRAFT DRAFT DRAFT DRAFT
\begin{abstract}
Using a technique that is not subject to first-order  \SU  symmetry breaking effects, 
we determine the $V_{us}$ element of the CKM matrix from data on semileptonic hyperon decays. 
We obtain $V_{us}=0.2250 (27)$, where the quoted uncertainty is purely experimental. This value 
is of similar experimental precision to the 
one derived from $K_{l3}$, but it is higher and thus in better agreement with the unitarity 
requirement,  $|V_{ud}|^2+|V_{us}|^2+|V_{ub}|^2=1$. An overall fit including the axial contributions, and neglecting \SU\!\! breaking corrections, yields  $F + D = 1.2670 \,\pm 0.0035$ and 
$F - D = -0.341 \pm 0.016$ with $\chi^{2} = 
2.96/3$ d.f.
%Valid PACS numbers may be entered using the \verb+\pacs{#1}+ command.
\end{abstract}

% PACS, the Physics and Astronomy
                             % Classification Scheme.
%\keywords{Suggested keywords}%Use showkeys class option if keyword
                              %display desired
\maketitle

%\section{\label{sec:level1}Introduction\protect\\} 

The determination of  the elements of the Cabibbo-Kobayashi-Maskawa (CKM) matrix \cite{cab,km}
is one of the main
ingredients for evaluating the solidity of the standard model of elementary particles.
This is a vast subject which has seen important progress with the  determination 
\cite{KTeVepsi, Lai:2001ki} of 
$\epsilon'/\epsilon$ and the observation \cite{Babar, Belle} of $CP$ violation in B decays.

While a lot of attention has recently  been justly devoted to the higher mass sector of
the CKM matrix, it is the low mass sector, in particular $V_{ud}$ and $V_{us}$ where 
the highest precision  can be attained. The most sensitive test 
of the unitarity of the CKM matrix is provided by the relation 
$|V_{ud}|^2+|V_{us}|^2+|V_{ub}|^2=1 - \Delta$. Clearly the unitarity condition is $\Delta = 0$.
The $|V_{ub}|^2$ contribution \cite{Vub} is negligible ($~10^{-5}$) at the 
current level of precision.  
%the unitarity test reduces to the consistency of 
%$\cos \theta_C$ determined from nuclear beta decay and  of  $\sin \theta_C$ 
%determined from strangeness changing semileptonic decays.
The value $V_{ud}$ = 0.9740 \ensuremath{\pm} 0.0005 is obtained from 
superallowed pure Fermi nuclear decays \cite{Towner}.
In combination with $V_{us}$ = 0.2196 \ensuremath{\pm} 0.0023, derived from $K_{e3}$ decay 
\cite{Leutwyler,PDG2002},
this yields $\Delta = 0.0032 \ensuremath{\pm} 0.0014$. On its face, this represents a 2.3 
standard deviation departure from unitarity \cite{Towner}.

In this communication we reconsider the contribution that the hyperon beta 
decays can give to the determination of  $V_{us}$. The conventional analysis
of hyperon beta decay  in terms of the parameters $F, D$ and $V_{us}$ is marred
by the expectation of first order $SU(3)$ breaking effects in the axial-vector
contribution. The situation is only made worse if one introduces adjustable 
$SU(3)$ breaking parameters as this  increases the number of degrees of freedom and
degrades the precision. If on the contrary, as we do here, one focuses the analysis 
on the vector form factors, treating the rates and $g_1/f_1$ \cite{GarciaBook} as the basic experimental data, 
one has directly access to the $f_1$  form factor  for each decay, and this in turn allows 
for a redundant determination of $V_{us}$. The consistency of the values of  $V_{us}$ 
determined from the different decays is a first confirmation of the overall consistency
of the model.
%****
A more detailed version of this work will be published in the Annual Reviews of Nuclear and Particle Sciences
\cite{CSW}.

In 1964 Ademollo and Gatto proved \cite{ad} that  
there is no first-order correction to the vector form factor,
$\Delta^1 f_1(0)=0$. 
This is an important result: since experiments can measure 
$V_{us} f_1(0)$,  knowing the value of  $f_1(0)$ in 
$\Delta S =1 $ decays is essential for determining $V_{us}$.
%This suggests a precise strategy in analyzing experiments to extract $V_{us}$. 
%We will use the information available from rates and angular correlations 
%to extract in each decay the value of $V_{us} f_1(0)$. 
%The Ademollo-Gatto theorem guarantees that we can compute the value of 
%$ f_1(0)$ with a satisfactory precision. Each decay thus provides 
%a value for $V_{us}$. If the theory is correct, these should coincide within errors,
%and can be fitted to obtain a best value of $V_{us}$.
%\section{Cabibbo-Model Fits}

The Ademollo-Gatto Theorem suggests an analytic approach to the 
available data that first examines the vector form factor $f_{1}$ 
because it is not subject to first-order  \SU  symmetry breaking 
effects. An elegant way to do this is to use the \emph{measured} 
value of $g_{1}/f_{1}$ along with the predicted values of $f_{1}$ 
and $f_{2}$ to extract a $V_{us}$ value from the decay rate 
for each decay. If the theory is correct, these should coincide within errors,
and can be combined to obtain a best value of $V_{us}$.
This consistency of the $V_{us}$ values obtained from 
different decays then indicates the success of the Cabibbo model. A
similar approach appears to have been taken in Ref. \cite{GV}.

Four hyperon beta decays have sufficient data to perform this 
analysis: $\Lambda \rightarrow p\,e^-\bar\nu,\; 
 \Sigma^- \rightarrow n\,e^-\bar\nu,\; 
\Xi^-\rightarrow \Lambda\,e^-\bar\nu,\;  
\Xi^{0}\rightarrow \Sigma^+\,e^-\bar\nu$ \cite{PDG2002}. 
Table \ref{Table:Vus} shows the results for them. 
In this analysis, both model-independent and model-dependent radiative 
corrections \cite{GarciaBook} are applied and q$^2$ variation of f$_{1}$ 
and g$_{1}$ is included.  Also \SU values of g$_{2}$ = 0 
and f$_{2}$ are used along with the numerical rate 
expressions tabulated in Ref. \cite{GarciaBook}.  We have not however included
 \SU\!\!-breaking corrections to the f$_{1}$ form factor, which will be discussed
 in the next section. The stated $V_{us}$ errors are purely experimental, coming
from experimental uncertainties in the hyperon lifetimes, branching ratios, 
and form factor ratios.  

The four values are clearly consistent 
$(\chi^{2} 
= 2.26/3$d.f.) with the combined value of $V_{us}$ = 0.2250 
\ensuremath{\pm} 
0.0027. This value is nearly as precise as that obtained from 
kaon decay ($V_{us}$ = 0.2196 \ensuremath{\pm} 0.0023) and, as 
observed 
in previous analyses \cite{Gaillard:1984ny, highvus, 
Flores-Mendieta:1998ii}, is somewhat larger. In combination 
with $V_{ud}$ = 0.9740 \ensuremath{\pm} 0.0005 obtained from 
superallowed 
pure Fermi nuclear decays \cite{Towner}, the larger $V_{us}$ value from  hyperon decays beautifully satisfies the unitarity constraint 
{\textbar}$V_{ud}${\textbar}$^{2}$ + 
{\textbar}$V_{us}${\textbar}$^{2}$ + 
{\textbar}$V_{ub}${\textbar}$^{2}$ 
= 1. 
% This is the Vus Analysis Table
\begin{table}[htb]
\begin{center}
\caption{Results from $V_{us}$ analysis using measured $g_1/f_1$ values}
\begin{tabular}{@{}lccc@{}}
\hline
\hline
Decay          &        Rate        &   $g_1/f_1$   &   $V_{us}$   \\  
Process        &  ($\mu sec^{-1}$)  &               &       \\  
\hline
$\Lambda \rightarrow p e^- \overline{\nu}$ &  $3.161(58)$      &  $0.718(15)$  &  0.2224 $\pm$ 0.0034  \\
$\Sigma^- \rightarrow n e^- \overline{\nu}$ & $6.88(24)$       &  $-0.340(17)$ &  0.2282 $\pm$ 0.0049   \\
$\Xi^- \rightarrow \Lambda e^- \overline{\nu}$ & $3.44(19)$    &  $0.25(5)$     &  0.2367 $\pm$ 0.0099  \\
$\Xi^0 \rightarrow \Sigma^+ e^- \overline{\nu}$ & $0.876(71)$  &  $1.32(+.22/-.18)$  &  0.209 $\pm$ 0.027  \\
Combined   &    ---    &     ---     &     0.2250 $\pm$ 0.0027 \\
\hline
\end{tabular}
\label{Table:Vus}
\end{center}
\end{table}
%\section{Possible Effects of \SU breaking}

We will limit our discussion to the effects that are most relevant for the determination of $V_{us}$.
Turning our attention first to \SU\!\!-breaking corrections to the $f_1$ form factor, we find in the literature 
computations that use some version of the quark model, as in \cite{Donoghue:th,Schlumpf:1994fb},
or some version of chiral perturbation theory, as in \cite{Krause:xc, 
Anderson:1993as, Flores-Mendieta:1998ii}. 

The quark-model 
computations find that the $f_1$ form factors
for the different $\Delta S=1$ decays are  reduced by a factor, 
the same for all decays, given as $0.987$  in \cite{Donoghue:th},  and $0.975$ 
in \cite{Schlumpf:1994fb},  a decrease respectively of $1.3\%$ or $2.5\%$.
This is a very reasonable result, the decrease arising from the 
mismatch of the wave functions of baryons containing different numbers
of the heavier $s$ quarks.

Evaluations of  $f_1$ in chiral perturbation theory  range from 
small negative corrections  in  \cite{Krause:xc} to larger
positive corrections in \cite{Anderson:1993as, Flores-Mendieta:1998ii}.
Positive corrections in $f_1$ for \emph{all} hyperon  beta decays 
cannot be excluded, but are certainly not expected in view of an argument \cite{Quinn:1968qy} 
according to which one expects a negative correction to $f_1$ at least in the 
$\Sigma^{-}\rightarrow n \, e^{-} \bar\nu$ case.
This result follows from the observation that the intermediate states that contribute
to the positive second-order terms in the Ademollo and Gatto sum rule have, in this case,
 quantum numbers 
$S=-2 ,\; I = 3/2$; no resonant baryonic state is known with these quantum numbers.
If we accept the hypothesis that the contribution of resonant hadronic states dominate, 
we can conclude that the correction to $f_1$ in $\Sigma^-$ beta decay 
should be  negative. 
We note that this argument also 
applies  to $K_{l 3}$ decays, and that the corrections
to these decays, computed with chiral perturbation theory,
are, as expected, negative. 

A modern revisitation of the quark-model computations 
will be feasible in the near future with the technologies of lattice QCD, and we would 
expect that a small negative correction would be obtained
in \emph{quenched} lattice QCD, an approximation that 
consists in neglecting components in the wave function of the baryons 
with extra quark-antiquark pairs. This is known to be an excellent
approximation in low-energy hadron phenomenology \cite{quenched}.

Multiquark effects can be included in  lattice  QCD by forsaking the 
quenched approximation for a  \emph{full} simulation. Alternatively one could 
resort to chiral perturbation theory to capture the major part of the multiquark
contributions which will be dominated by virtual $\pi,\, K,\,\eta$ states. Early results of a
similar strategy applied to the $K_{e3}$ decays \cite{Martinelli} indicate that in that case a
1\% determination of the $f_{+}(0)$ form factor is within reach, and we expect that
a similar precision can be obtained in the case of hyperon decays. 
In the present situation we consider it best not to include any \SU\!\! breaking corrections in
our evaluation, nor to include an evaluation of a theoretical error. Our expectation
that the corrections to  $f_{+}(0)$ will be small and negative can only be substantiated 
by further work.
 
We next turn our attention to the possible effect of ignoring the $g_2$ form factor.
In the absence of second class currents \cite{WeinbergII}
the form factor $g_2$ can be seen to vanish in the 
\SU  symmetry limit. The argument is very straightforward: the 
neutral currents $A^3_\alpha= \bar q \lambda^3 \gamma_\alpha \gamma_5 q$
and $A^8_\alpha= \bar q \lambda^8 \gamma_\alpha \gamma_5 q$
that belong to the same octet as the weak 
axial current are even under charge conjugation, so that their 
matrix elements cannot contain a weak -- electricity term, which is $C$-odd.
The vanishing of the weak electricity in the proton
and neutron matrix elements of  $A^3_\alpha,\, A^8_\alpha$ implies 
the vanishing of the $D$ and $F$ coefficients for $g_2(0)$, 
so that, in the \SU limit, the $g_2(0)$ form factor vanishes for any current in the octet. 

In hyperon decays  a nonvanishing $g_2(0)$ form factor can arise from the 
breaking of \SU  symmetry. Theoretical estimates \cite{Holstein} 
indicate a value for $g_2(0)/g_1(0)$ in the $-0.2$ to 
$-0.5$ range.

In determining the axial-vector form factor $g_1$ from the Dalitz Plot --- or, equivalently, the electron--neutrino correlation --- one is actually measuring  
$\tilde g_1$, a linear combination of $g_1$  and $g_2$  
($\tilde g_1 \approx g_1 - \delta g_2$   up to first order in 
$\delta = \Delta M/M$). This has already been noticed in past experiments 
and is well 
summarized in Gaillard and Sauvage \cite{Gaillard:1984ny}, Table 8. 
Therefore, in deriving $V_{us}^2f_1^2$ (hence $V_{us}$) from the beta decay rate, there is in fact a small sensitivity to  $g_2$. 
To first order,  the rate is proportional to   $V_{us}^2[f_1^2 + 
3 g_1^2 - 4 \delta\, g_1 g_2] \approx V_{us}^2[f_1^2 + 3 \tilde g_1^2 +2 \delta\, \tilde g_1 g_2]$. In fact, this is a second
order correction to the value of $V_{us}$, potentially of the same order of magnitude
as the corrections to $f_1$.

Experiments that measure correlations with polarization --- in addition to the electron--neutrino 
correlation --- are sensitive to $g_2$. While the data are not yet sufficiently precise to yield good quantitative information, one may nevertheless look for trends. In polarized  $\Sigma^-\rightarrow p\,e^-\bar\nu$  \cite{E715}, 
negative values of $g_2/f_1$ are clearly 
disfavored (a positive value is preferred by $1.5\sigma$). Since the same experiment unambiguously established that $g_1/f_1$  is negative one concludes that allowing for nonvanishing $g_2$  would increase the derived value of  $V_{us}^2f_1^2$. 
In polarized  $\Lambda \rightarrow p\,e^-\bar\nu$ the data favor \cite{Oehme} negative values of  $g_2/f_1$  (by about $2\sigma$).  In this decay, $g_1/f_1$  is positive so that again, allowing for the presence of nonvanishing $g_2$  would increase the derived value of $V_{us}^2f_1^2$. In either case, 
we may conclude that making the conventional assumption of 
neglecting the $g_2$ form factor tends to \emph{underestimate} the derived 
value of $V_{us}$. 
A more quantitative conclusion must await more precise experiments. We consider
it to be of the highest priority to determine the $g_2$ form-factor (or a stringent limit
on its value) in at least one of the hyperon decays, ideally in $\Lambda$ semileptonic 
decay which at  the moment seems to offer the single most precise determination of $V_{us}$ .

The excellent agreement with the unitarity condition of our determination 
of $V_{us}$, which neglects \SU\!\!-breaking effects, seems to indicate that such
effects were overestimated in the past, probably as a consequence of  the uncertainties of the
early experimental results.
We also find \cite{CSW} that the $g_1$ form factor of the different decays,
which is subject to first order corrections, is well fitted by the $F, D$ parameters 
\cite{cab}, with  $F + D = 1.2670 \,\pm 0.0035$ and 
$F - D = -0.341 \pm 0.016$ with $\chi^{2} = 
2.96/3$ d.f.
%\section{Conclusions}

The value of $V_{us}$ obtained from hyperon decays is of comparable experimental 
precision with that obtained from $K_{l3}$ decays, and is in better agreement
with the value of \Tc obtained from nuclear beta decay. While a discrepancy 
between $V_{us}$ and $V_{ud}$ could be seen as a portent of exciting new physics, 
a discrepancy between the two different determinations of   $V_{us}$ can only be 
taken as an indication that more work remains to be done both on the theoretical and
the experimental side. 

On the theoretical side, renewed efforts are needed for the determination 
of  \SU\!\!-breaking effects in hyperon beta decays as well as in $K_{l 3}$ 
decays. While it is 
quite possible to improve the present situation on the quark-model 
front, the best hopes lie in  lattice 
QCD simulations, perhaps combined with chiral perturbation theory
for the evaluation of large-distance multiquark contributions.

We have given some indication that  the trouble could arise from the 
$K_{l 3}$ determination of $V_{us}$, and we would like to encourage further experimental 
work in this field \cite{newk}. 
We are however convinced of  the importance of
renewed experimental work on hyperon decays, of the kind now in progress
at the CERN SPS. The interest of this work goes beyond  the determination of
$V_{us}$, as it involves  the intricate and elegant relationships that the
model predicts. 

The continuing intellectual stimulation provided by colleagues in the Fermilab KTeV Collaboration, particularly members 
of the hyperon working group, is gratefully acknowledged. This work was supported in part by the U.S. Department of Energy under grant 
DE-FG02-90ER40560 (Task B).

\end{document}